\documentclass[11pt]{article}
\begin{document}

\title{THE EFFECTS OF THE ROTATION IN PLASMA}
\author{S. N. Arteha}
\date{\small Space Research Institute, Profsoyuznaya 84/ 32, Moscow 117810, Russia}
\maketitle

\begin{abstract}
The electric and magnetic self-fields can exist in the rotating plasma.
The self-sustained rotation can be established in the plasma.
The disturbed distribution
function of rotating plasma is derived from the Vlasov
equation. The propagation of waves
in rotating plasma differs from that in the usual plasma. New terms
of the Landau damping appear. The local rotational behaviour may become
prevailing.
\end{abstract}

\section{INTRODUCTION}

Phenomena dealt with are not too rare. However, the causes
of the phenomena (forces, fields) are rather weak and can often
be supposed to be negligible.

It is well known that many astronomical and laboratory objects represent
rotating plasma. Therefore, the study of
rotating plasma is of great theoretical and practical interest.

It is assumed that plasma is neutral on the average.
However, electric and magnetic
fields can exist in plasma, because there are forces
which oppose the electric one$^{1,2}$.

It is well known that in a wide variety of plasma devices
plasma travels across the magnetic field at rates much higher
than those explained by collisional processes. It has been
assumed that the electron transport across the magnetic field
arises from the $\vec{E}\times\vec{B}$ drift
($\vec{E}\times\vec{B}$ instability$^3$).

The radial electric field exists to balance out the
$\vec{v}\times\vec{B}$ and centrifugal force terms$^4$.

The plane sheet stability
of a surface charge has been considered for externally applied and
mutually perpendicular fields ($\vec{E}\perp\vec{B}$) which
produce an equilibrium particle flow in the sheet$^5$.
The calculation was extended to a cylindrical shell in the
inhomogeneous magnetic field$^6$
(the size of a shell is identical to the particle gyro-orbit).

The equilibrium properties
induce a self-electric field for a charge slab or a cylinder (the fluid
approximation$^{7,8}$).

The effect of equilibrium radial electric field on the trapped - particle
stability has been investigated $^9$. The stabilization can result
from the configuration in which the radial potential assumes a hollow
profile (the equilibrium radial electric field points inward). The
theory also predicts the possibility of additional instability, when
the $\vec{E}\times\vec{B}$ rotation is present (the rotationally driven
trapped - particle mode).

The development of a radial
electric field $E_\rho$ due to toroidal rotation has been
studied for the toroidal geometry $^{10}$.
The ion distribution function becomes
Maxwellian with temperature uniform on the magnetic surface.
The poloidal flow decays. The density is distributed
over magnetic flux surfaces by the Boltzmann factor with the
effective potential energy that is the sum of a centrifugal
potential and an electrostatic potential required for charge
neutrality. The usual small gyroradius
expansion of  the Fokker-Planck equation was carried out,  and it
was shown, that the effects of centrifugal and Coriolis forces on
the particle motion affect the transport.

The rotational equilibrium in cylindrical, non-neutral
plasma with an axial magnetic field has been analysed in
detail $^{11}$.

The poloidal rotation of plasma due to the
$\vec{E}\times\vec{B}$ drift has been studied in
tokamaks (the so-called $H-mode$ of improved
confinement) with a radial electric field and strong axial
magnetic field. The $\vec{E}\times\vec{B}$ drift rotates the
plasma poloidally (when the radial electric field exists). In
non-$H$-mode plasmas the poloidal rotation frequency is
typically much less than the toroidal rotation frequency in
tokamaks. The toroidal geometry of tokamaks becomes important
in some theories of poloidal spin-up$^{12}$.
The poloidal rotation instability (resulting from the
poloidally asymmetric diffusion) leads to poloidal
velocity shears which may quell the microturbulence. The nonlinear
interplay between the poloidal spin-up and turbulence-driven
anomalous transport leads to bifurcated equilibria (the $L$ to
$H$ transition mode in tokamaks). The tokamaks can spontaneously
develop a poloidal velocity shear (the "P-S" flows). The spin-up
persists even under conditions where the growth rate is sonic $^{13}$.

The rotation is well known to induce new poles in the
perturbed distribution function. The so-called resonant
diochotron instability $^{14,15}$ involves the Landau damping at
the resonant frequency $\omega=I\Omega$ ($I$ is an
integer, $\Omega$ is the rotation frequency). The role of
Landau damping in cross-field electron beams and in an inviscid
shear flow has been considered for the cylindrical geometry
(perturbations with $k_z=0$) with unperturbed angular velocity
$\omega_0(r)={1\over 2}n_{0}, 0\le r < b$ and
$\omega_0(r)={1\over 2}n_{0}(b/r)^2, b <r \le a$ 
($n_{0}$ is a constant). A formal analysis of the
resonance has been given in the Laplace
transforms context $^{14}$. The case of a perfectly
conducting wall (at the impedance $Z(\omega)=0$) was studied, and
a slight shift in the eigenfrequencies from their
locations for an induced $Z(\omega)\ne 0$ was found. The Landau damping
can overcome the resistive destabilization (if there exist
gradients in the electron density) for a strictly decreasing
electron density in a cylindrical system, so that only one resistively
unstable normal mode can exist - the fundamental one.

The diochotron instability of a thin (${r_2-r_1\over r_1} \ll 1$), tenuous
($\omega_p \ll \omega_{Be}$), cylindrical
layer (a single-component plasma)
has been investigated in a strong uniform constant magnetic
field using the Vlasov equation $^{15}$.
The particle gyro-radius was of the order of the mean radius of a layer.
The theoretical
model does resemble the geometry of some experiments which
rely on the high energy particles injection.
Two distribution functions were considered,
which give the same density ($\rho (r) \sim [\Theta
(\vec{r}-\vec{r}_1) - \Theta (\vec{r}-\vec{r}_2)]$, where
$\Theta$ is the Heaviside step function), but are different in
the velocity space (the velocity spread is zero, or the
particles oscillate around the mean radius). The self-electric
field of the equilibrium density (which is strictly radial) was
included in the treatment (but not the self-magnetic
field). The stability against  electrostatic  perturbations  with
frequency around  $-l\omega_{Be}$  ~(where  $l$  is the asimuthal
wave number of perturbations) and the infinite wavelength in  the
$z$-direction has been investigated. The perturbed
distribution function was found, and the macroscopic sheared flow
was shown to be not a relevant criterion for instability by itself.

The gyrokinetic integral equations, used for studying the ion temperature
gradient (ITG)- driven mode in a toroidal geometry at low plasma pressure,
have been extended to include the equilibrium ion sheared flows $^{16}$. The
parallel and perpendicular sheared flows are destabilizing and stabilizing
mechanisms, respectively. The gyrokinetic integral code was used to explore the
sheared flows' effects on the ITG mode in a sheared slab geometry $^{17}$. The
mode structure and eigenfrequencies predicted by the integral code differ
from the results derived by the differential approach (for higher radial
eigenmodes).

The further development of the above mentioned
works and some new ideas are presented in this work. The
purposes of the paper are: \\
1) to show the separation of charges and to obtain rigorous
expressions for electric and magnetic self-fields in rotating
plasma;\\
2) to demonstrate the self-sustained rotation origin;\\
3) to derive the perturbed distribution function and to find
the location of poles (the terms in the Landau
damping);\\
4) to propose the local-rotational description of
plasma.

The consideration is self-consistent in its character (from the first
principles) and leans upon the microscopic properties of a
matter, rather than upon the continuous medium (fluid) models.
The description can be applied to free rotating plasma objects
(astronomical, for  example)  with  some   arbitrary  shape   and
composition, to  plasma  devices  with  cylindrical  symmetry  in
particular (the arbitrary geometry is
included as the limits of integration only).

The peculiarities, which differentiate this work from the
cited papers, are as follows: \\
- Not only the radial electric self-field $E_{\rho}(\rho,z)$,
but the magnetic self-field $B_z(\rho,z)$ and
smaller self-fields $E_z(\rho,z)$ and
$B_{\rho}(\rho,z)$ are taken into consideration  (note  that  all
these components are functions not only of $\rho$ but of $z$ as well).
This consideration is necessary, since the behaviour of particles changes 
qualitatively: the solution depends not only on the small ratio 
${v\over c}$, but on arbitrary ratio of characteristic velocities as well. 
All quantities (density, charge density, local composition, etc.) are not externally given but
can be evaluated in a self-consistent manner. \\
- The possibility of a steady-state rotation of
plasma is shown. The self-sustained rotation of initially neutral
plasma origins in the external magnetic field $\vec{B}_0$ as a
consequence of the $\vec{E}\times\vec{B}$ drift ($\vec{E}$ is
the self-field).\\
- The perturbed distribution function is found (rigorously)
proceeding from the Vlasov equation. The approach differs in
carrying through the extra terms due to $k_z$, a
self-consistent radial centrifugal force, the self-fields and
in allowing the wave amplitudes to be space-dependent.\\
- The new poles are shown to exist, and the location of these
poles is found (the new terms in the Landau damping
appear). Since the wave propagation in rotating plasma (an
accelerated medium) differs from that in ordinary plasma, the
local-rotational description of plasma is supposed to be possible
(in the strong turbulence case, for example).

\section{THE ELECTRIC AND MAGNETIC FIELDS IN ROTATING PLASMA}

Rotating plasma possesses some peculiarities as compared to rotating objects
which consist of neutral particles. We start with the rigid rotation for
a simple explanation only (the substitutions for an arbitrary rotation
are given below). The existence of the centrifugal force
\begin{equation}
\vec{F} = m\Omega^{2}\vec{\rho},
\end{equation}
(where $\vec{\Omega}$ is the angular frequency of the system,
$m$ is the particle mass, $\vec{\rho}$ is its distance from
the axis of rotation) causes different effects on particles of 
different masses. As a result of
these different effects, the $\rho$-dependence of particle concentration
is bound to be different for particles of different masses.

According to the Boltzmann distribution,
\begin{equation}
n_{\alpha} = n_{0\alpha}\exp{\Biggl ( {-U_{\alpha}\over kT}\Biggr ) },
\end{equation}
where $T$ is the system temperature, $k$ is the Boltzmann constant,
$n_{0\alpha}$ is the particle concentration on the axis of rotation, $U_{\alpha}$
is the potential energy of $\alpha$-sort particles. In the
case of neutral particles
\begin{equation}
U_{\alpha} = -{m_{\alpha}\Omega^{2}\rho^{2}\over 2}.
\end{equation}
The $n_{\alpha}(\rho)$ dependences for hydrogen plasma
are shown schematically in Fig.1. Here $R_{0}$ is the size of a system
(the distance from the axis of rotation); symbol $\alpha$ is
either $H$ (hydrogen) or $e$ (electron).

However, plasma consists of charged particles. Since the particles of different
masses in plasma have different charges, this partial separation of
particles produces a partial separation of charges. The
negatively charged region is bound to lie near the axis of rotation, whereas the
positively charged region lies near the system
boundary ($R_{0}$). The electric field $\vec{E}_{0}(\rho,z)$
(in the polar coordinate system: the $\vec{\Omega}$ direction is the z direction)
exists as a result of charge separation.
This field opposes the considerable separation of charges.

\begin{figure}
\unitlength=1mm
\special{em:linewidth 0.4pt}
\linethickness{0.4pt}
\begin{picture}(115.00,90.00)
\put(5.00,20.00){\vector(0,1){65}}
\put(5.00,20.00){\vector(1,0){75}}
\put(0.00,15.00){$0$}\put(0.00,80.00){$n_{\alpha}$}\put(80.00,15.00){$\rho$}
\put(68.00,15.00){$R_0$}\put(50.00,15.00){$R'_1$}\put(65.00,77.00){$n_i$}
\put(65.00,45.00){$n_e$}\put(20.00,5.00){FIG.1. The dependences $n_{\alpha}(\rho)$.} 
\multiput(70.00,20.00)(0.00,5.00){12}{:}
\multiput(52.00,20.00)(0.00,5.00){5}{:}
\qbezier(5.00,26.00)(67.00,35.00)(70.50,80.00)
\qbezier(5.00,30.00)(63.00,40.00)(70.50,62.00)
\end{picture}
\end{figure}

As illustrated in Fig.1, there is some distance from the axis of rotation
$R'_{1}(z)$, where the local charge equals zero. For $0 \leq \rho < R'_{1}(z)$
the plasma is negatively charged on the average;
for $R'_{1}(z) < \rho \leq R_{0}$ the plasma is
positively charged.

The partial separation of charges gives rise to
the $\rho$-dependence of charge density; that is, the plasma
possessing the given charge density moves round a circle of definite radius.
Therefore, circulating currents are inside the rotating plasma despite
the fact, that all particles revolve with $\vec{\Omega}$ on the average
($\vec{\Omega}$ has a distinct direction). For $0 < \rho < R'_{1}$
the current is opposite to the plasma rotation direction (since
the charge is negative), whereas for $R'_{1} < \rho \leq R_{0}$ the current
flows in the direction of rotation (since the charge is positive).
In the general case the magnetic actions of currents are not compensated,
and the magnetic field $\vec{B}$ exists. This is the magnetic field of a solenoid, whose
axis coincides with the axis of system rotation.

For rotating, fully ionized plasma (near macroscopic
equilibrium), which consists of elements with
atomic number $N_{i}$, it follows in polar coordinates $\rho, \varphi, z$
($z$ axis coincides with the $\vec{\Omega}$ direction), that
$n_{e}, n_{i}$ can be taken from (2) with
$$
U_{e} = -{m\Omega^{2}\rho_{1}^{2}\over 2} +
e\int_{0}^{\rho_{1}} E_{0\rho}(\rho',z_{1})d\rho' +
e\int_{0}^{z_{1}} E_{0z}(\rho_{1},z')dz' +
$$
\begin{equation}
\int_{0}^{\rho_{1}} {eB_z\over c}\Omega\rho'd\rho' -
\int_{0}^{z_{1}} {eB_{\rho}\over c}\Omega\rho_1dz'~ ,
\end{equation}
$$
U_{i} = -{M\Omega^{2}\rho_{1}^{2}\over 2}
- eN_{i}\int_{0}^{\rho_{1}}
E_{0\rho}(\rho',z_{1})d\rho' - eN_{i}\int_{0}^{z_{1}} E_{0z}(\rho_{1},z')dz' -
$$
\begin{equation}
\int_{0}^{\rho_{1}} {eB_z\over c}\Omega\rho'd\rho' +
\int_{0}^{z_{1}} {eB_{\rho}\over c}\Omega\rho_1dz'~ ,
\end{equation}
where $n_{e}(\rho,z)$ is the electron concentration, $n_{i}(\rho,z)$ is the
ion concentration, $-e$ is the electron charge ($e>0$), $m$ is the
electron mass, $M$ is the ion mass, $T_{e} = T_{i} = T$ are
electron and ion temperatures, $\vec{r}$ is the radius-vector: $\vec{r} = \rho\vec{e_{\rho}} + z\vec{e_{z}}$.
The quantities $n_{0\alpha}$ can be found from conditions: ~ $\int_{(V)} n_{\alpha}dv' = N_{\alpha}K_{\alpha},$ where
$K_{\alpha}$ is the total number of $\alpha$-sort particles,
$N_e=1$ and $K_{e} = N_{i}K_{i} + \Delta$, $\Delta$ is
the electron surplus (for non-neutral plasma $\Delta \ne 0$).

The $\vec{E}_{0}(\rho,z)$ components can be obtained from the
system of integral equations:
\begin{equation}
E_{0\rho}(\rho,z) = e\int_{(V)} {q(\rho_{1},
z_{1})\over R_1^3}\rho_{1}[\rho-\rho_1cos(\varphi_1-\varphi)]d\rho_{1}dz_{1}d\varphi_1
~,
\end{equation}
where
\begin{equation}
q(\rho_{1}, z_{1}) =
n_{0i}\exp{\Biggl ( -{1\over kT}U_{i}\Biggr ) } - n_{0e}\exp{\Biggl ( -{1\over kT}U_{e}\Biggr ) },
\end{equation}
$$
R_1 = \sqrt{(z-z_1)^2+\rho^2+\rho_1^2-2\rho\rho_1cos(\varphi_1-\varphi)}
$$
\begin{equation}
E_{0z}(\rho,z) = e\int_{(V)} {q(\rho_{1}, z_{1})\over R_1^3}(z-z_1)\rho_{1}d\rho_{1}dz_{1}d\varphi_1~ ,
\end{equation}
$V$ is the system volume (its shape is arbitrary).

Radius $R'_{1}$ can be found from the equation $n_{e}(R'_{1}) = n_{i}(R'_{1})$
using (2), (4)-(8).

It follows from the Biot-Savart law, that the magnetic field is
$$
\vec{B} = {1\over c}\int_{(V)} {[\vec{j_1} \times
\vec{R_1}]\over R_1^{3}}dv_1 ~,
$$
where $c$ is the speed of light, $dv_1$ is the region with currents, $R_1$ is the
distance of this region to the point of observation, $\vec{j_1}$ is the
current density.

The magnetic field due to rotation can be obtained from the following expressions:
\begin{equation}
B_{\rho} = {1\over c}\int_{(V)}
{j_{\varphi}(\rho_1,z_1)\over R_1^3}(z-z_1)\rho_1d\rho_1dz_1d\varphi_1~ ,
\end{equation}
\begin{equation}
B_{z} = {1\over c}\int_{(V)}
{j_{\varphi}(\rho_{1},z_{1})\over
R_1^3}\rho_{1}[\rho_1cos(\varphi_1-\varphi)-\rho]d\rho_{1}dz_{1}d\varphi_1
~,
\end{equation}
where
\begin{equation}
j_{\varphi}(\rho_{1},z_{1}) = e\Omega\rho_{1}q(\rho_1, z_1) ,
\end{equation}
with substitutions $q(\rho_1, z_1)$ from (7), $U_{\alpha}$ from (4),
(5) and $\vec{E}_{0}(\rho',z')$ from (6), (8); the magnetic field near the axis of
rotation at the system boundary is $B_z$ from (10) for $z=L,
\rho =0$, where $2L$ is the system size along the axis of rotation.

The following substitutions need to be done
for rotating plasma of complex composition: \\
$j_{\varphi} \rightarrow \sum_{l} j_{\varphi}^{l}~ ,~ n_{0\alpha} \rightarrow n_{0\alpha}^{l};$
in the case of unsteady rotation
$$
{\Omega^{2}\rho^{2}\over 2} \rightarrow
\int_{0}^{\rho} \Omega(\rho', z)\rho'd\rho'.
$$

For astronomical objects the effects can be considerable; in this case
the gravitational force plays important role. Note, that in
the general case the field $E_{\varphi}\vec{e}_{\varphi}$
exists for free rotating objects with
some arbitrary (non-symmetric) shape. In this case all quantities
depend on coordinate $\varphi$ as well. The additional terms
$$
e\int_0^{\varphi} E_{0\varphi}\rho_1d\varphi_1 ~, ~~~~
-eN_i\int_0^{\varphi} E_{0\varphi}\rho_1d\varphi_1
$$
appear in (4) and (5), respectively. The component
$E_{0\varphi}$ is taken from
$$
E_{0\varphi}(\rho, z, \varphi) = e\int_{(V)} {q(\rho_{1},
z_{1}, \varphi_1)\over
R_1^3}\rho_{1}^2sin(\varphi-\varphi_1)d\rho_1dz_1d\varphi_1 .
$$

\section{SOME APPROXIMATION OF THE DIELECTRIC CONSTANT TENSOR}

Now we obtain information about the electron part of the
dielectric constant tensor $\varepsilon_{\alpha\beta}$ (the ion part
can be found in a similar manner). The polarization vector $\vec{P}$
is defined as
$$
{\partial\vec{P}\over \partial t} = \vec{j} ~, ~~~ \vec{P} =
\vec{P}_0 + \vec{P}'\exp{[i(\vec{k}\vec{r}-\omega t)]} .
$$
It follows that
\begin{equation}
P_{\alpha}' = {\varepsilon_{\alpha\beta} - \delta_{\alpha\beta}\over
4\pi}E_{\beta}^, ~,
\end{equation}
where $-i\omega\vec{P}'=-eK_e\vec{v}$, $\delta_{\alpha\beta}$
is the Kroneker symbol.
The expression for $\varepsilon_{\alpha\beta}$ in a rotating
system can be derived directly from the equations of electron
motion (the field $\vec{E}$ contains the wave field
$\vec{E}'\sim e^{-i\omega t}$) with the force (exact)
\begin{equation}
\vec{F}_e = -e\vec{E} - {e\over c}[\vec{v}\times\vec{B}] +
m\Omega^2\rho\vec{e}_{\rho} + 2m[\vec{v}\times\vec{\Omega}] -
{e\over c}B_z\Omega\rho\vec{e}_{\rho} + {e\over c}\Omega\rho
B_{\rho}\vec{e}_z .
\end{equation}
We suppose that the constant field $\vec{B}$ contains all
components ($B_{0\rho},B_{0\varphi},B_{0z}$) and $\vec{v} \sim
e^{-i\omega t}$, i.e. space variations of $\vec{E}'$ are neglected.
Resolving the equation $\vec{v}=i\vec{F}_e/(m\omega)$ in
terms of $v_{\rho}, v_{\varphi}, v_z$ and extracting the
$\vec{E}'$-dependence only (the constant polarization $\vec{P}_0$
and the constant part $\varepsilon_{\alpha\beta}^{(0)}$ exist), one obtains
\begin{equation}
\varepsilon_{\alpha\alpha} = 1 -
{\Omega_e^2[\omega^2-(\omega_{Be}^{\alpha})^2]\over \omega^2[\omega^2-\omega_1^2]}~,
\end{equation}
where $\Omega_e=(4\pi K_ee^2/m)^{1/2}~ , \omega_1=\sqrt{(\omega_{Be}^{\rho})^2+(\omega_{Be}^{\varphi})^2+
(\omega_{Be}^z)^2}~,\\
\omega_{Be}^{\rho}=eB_{0\rho}/(mc),~
\omega_{Be}^{\varphi}=eB_{0\varphi}/(mc)$~, $\omega_{Be}^{z}=eB_{0z}/(mc) - 2\Omega$~,
\begin{equation}
\varepsilon_{\alpha\beta} = {\Omega_e^2(\omega_{Be}^{\alpha}\omega_{Be}^{\beta}+i\omega\omega_{Be}^{\gamma})\over \omega^2[\omega^2-\omega_1^2]}~,
\end{equation}
here $\alpha$, $\beta$ and $\gamma$ are taken from cyclic
rearrangements in $\rho, \varphi, z$;
$\varepsilon_{\alpha\beta}=\varepsilon_{\beta\alpha}^*$.
The approach requires the following conditions:\\
$v_T\mid k_{\alpha}\mid/\omega \ll 1~~, ~~~~ v_T\mid
k_{\beta}\mid/\omega_{Be}^{\alpha} \ll 1 ~, $
(the smallness of space $\vec{E}'$-variations in the localization
region of an electron), i.e. $\omega$ must be large enough;
and, besides, $\omega$ must not be near the frequency $\omega_1$.
It follows from the expressions (14), (15), that the
dissipation equals zero:
$$
Q = {i\omega\over 16\pi}(\varepsilon_{ik}^* -
\varepsilon_{ki})E_iE_k^* = 0
$$
(the tensor $\varepsilon_{\alpha\beta}$ is Hermitian). The
$\varepsilon_{\alpha\beta}$ values determine the wave polarization
(which is elliptic in the general case) for large
$\omega$. The gyration vector $\vec{G}$ can be found from the
equation $D_{\alpha} = \varepsilon_{\alpha\beta}E_{\beta} +
i[\vec{E}\times\vec{G}]_{\alpha}$ , here $i=\sqrt{-1}$.
In this approximation the components $G_{\alpha}$ are:
$$
G_{\alpha} = {\Omega_e^2\omega_{Be}^{\alpha}\over \omega [\omega^2-\omega_1^2]} ~,
$$
here $\alpha = \rho, \varphi, z$.

We note in the general case, that the space dependences of
fields $\vec{E},~\vec{B}$ and the wave amplitude need to be taken
into consideration, in principle (to obtain rigorous
expressions for $\varepsilon_{\alpha\beta}$
or damping of waves, for example). One may suppose, that, as
a consequence of space nonhomogeneity of all characteristics,
the rotating plasma system as a whole can be more stable with respect to
external disturbances (the system goes from resonance
influences). To impart some energy, the spectrum needs to be
rather wide-range.

\section{THE LOCATION OF POLES}

Since the  rotating  plasma is a nonhomogeneous accelerated medium,  the wave
propagation has some peculiarities (damping, non-rectilinearness).
In the general case the resonance conditions possess a relatively
complex structure $\psi (\rho,\varphi,z,\vec{B},\vec{E},\vec{v})=0$.
The derivation  of  the resonance conditions will be demonstrated below
in the context of the linear kinetic description for some special cases. 
Note, that the elementary  approach  to  the  problem  can  be
developed from the analysis of the exact force (13): 
it is necessary to find such
particles ($\vec{v}(\rho,\varphi,z,t)$), for which the work of the
wave field $A=\int \vec{F}\vec{v}dt$ is positive for some specific wave
(in type, direction, polarization). The value $A$ is
$$
A = \int \{-e(\vec{E}\vec{v}) + m\Omega^2\rho v_{\rho} +
{e\over c}\Omega\rho (B_{\rho}v_{z} - B_{z}v_{\rho})\}dt .
$$
If the mean work is positive,  the Landau damping mechanism takes
place.
In the kinetic description we start from the system of
self-consistent field equations:
\begin{equation}
{\partial f_{\alpha}\over \partial t} + \vec{v}_{\alpha}{\partial
f_{\alpha}\over \partial
\vec{r}} + \vec{F}_{\alpha}{\partial f_{\alpha}\over \partial \vec{p}} = 0 ,
\end{equation}
$$
rot \vec{E} = -{1\over c}{\partial\vec{B}\over \partial t}~ ,~~ div \vec{B} = 0 ,
$$
$$
rot \vec{B} = {1\over c}{\partial\vec{E}\over \partial t} + {4\pi\over c}\vec{j}~ ,~~ div \vec{E} = 4\pi q ,
$$
$$
q = e\int(N_if_{i} - f_{e})d^{3}p~ ,~~ \vec{j} = e\int(N_if_{i} -
f_{e})\vec{v}d^{3}p ,
$$
and separate the disturbed terms (caused by the wave
propagation):
\begin{equation}
f_{\alpha} = f_{0\alpha} + \delta f_{\alpha}~ ,~~~ \vec{B} =
\vec{B}_{0} + \vec{B}'~ ,~~ \vec{E} = \vec{E}_{0} + \vec{E}'~,
\end{equation}
where $\alpha$ means either $e$ or $i$.
It is believed that
\begin{equation}
\delta f_{\alpha},~ \vec{E}',~ \vec{B}' \sim A(\rho, z)\exp{i(\vec{k}\vec{r} - \omega t)} .
\end{equation}

The undisturbed distribution function (in the rotating
system) takes the form:
\begin{equation}
f_{0e} = {K_{e}D\over (2\pi mT)^{3/2}}\exp{\Biggl [ -{U'_{0}\over T}\Biggr ] },
\end{equation}
\begin{equation}
U'_{0} = {mv^{2}\over 2} + U_e,
\end{equation}
where $K_e$ is the total number of electrons in the volume $V$,
$D$ is the normalization constant,
$U_e$ is taken from (4). The expression (19) is a solution of the kinetic
equation (16) with the force (13). $U_e$ is the potential part of electron
energy (related with the force $\vec{F}$). Note that the nonconservative
force $e\vec{v}\times\vec{B}$ is not appeared in the potential at all.
All terms with $\vec{B}$ include the factor $\Omega$ and exist as a result
of the system rotation only.
We use the substitutions
$$
\vec{E}_{0} = E_{0\rho}(\rho,z)\vec{e}_{\rho} + E_{0z}(\rho,z)\vec{e}_{z},
$$
and ~~ $\vec{k'} = \vec{k} + S(\vec{r})\vec{r}$, where
\begin{equation}
S(\vec{r}) =    -{i\over    r^{2}}\Biggl (\vec{r}{\partial
\ln{A(\rho,z)}\over \partial\vec{r}}\Biggr ) .
\end{equation}

Now we make some remarks.\\
1) The accounting of the amplitude space-dependence
is equivalent to including some imaginary part in
$\omega$: ~$\omega \rightarrow \omega + i\omega'$. For, the
value $\omega'$ is greater than zero in the case of wave
damping, and it follows from (18), (21), that the appropriate
wave amplitude $A(\rho,z)$ is decreasing; the unstable case
$\omega'<0$ is analogous to wave amplitude increasing
(see (18),(21)). The collisional damping can also be included
in the term (21). The reverse problem can be formulated:
with the $A(\rho,z)$ dependence known (using a probe wave), one should
obtain information about collisional processes (the integral
of collisions). However, it is not our intention to follow
this way.\\
2) In the general case of unsteady rotation and gravitational
force existence the following substitutions need to be done:
$$
E_{0\rho} \rightarrow E_{0\rho1} = E_{0\rho} - {m_{e}\over e}g_{\rho}(\rho,z)~ ,~
E_{0z}^{(1)} \rightarrow    E_{0z1}^{(1)}    =   E_{0z}^{(1)}   -
{m_{e}\over e}g_{z}(\rho,z),
$$
in (20):
$$
{m\Omega^{2}\rho^{2}\over 2T}
\rightarrow {m\over T}\int_{0}^{\rho} \Omega^{2}(\rho_{1},z)\rho_{1}d\rho_{1} +
{m\over T}\int \vec{g}(\rho'z')d\vec{r}' ,
$$
where
$$
\vec{g}(\rho,z) = G{M(\rho)\vec{r}\over r^{3}} ,
$$
$G$ is the gravitational constant, $M(\rho)$ is the mass in the $0 \le \rho'
\le \rho$ region; in(13):
$$
m\Omega^{2}\rho\vec{e}_{\rho} \rightarrow m\Omega^{2}(\rho,z)\rho\vec{e}_{\rho} +
m\vec{g}(\rho,z).
$$
The derivation   of   the   disturbed  distribution  function  is
demonstrated in Appendix.

Of main interest here are the poles of $\delta f_{e}$ (they contribute to the
Landau damping). As a consequence of different regimes  of
particles behaviour  (see  Appendix),  the  resonance  conditions
differ depending on the particles velocity.
For $a < 1$ the term $Q(\varphi - \tau)$
does not have any pole
in an explicit form (see (A8) in Appendix). The imaginary part of the dielectric constant tensor arises
from the imaginary part of $\delta f_{M}$ and from the
integration of the exponent $^{18}$.
We consider the case $\mid a\mid \ll 1$, i.e. $v_{\perp}$ is
large enough:
$$
v_{\perp} \gg {\Omega_{1}^{2}\rho\over \omega_{Be}^{(z)}} ~,
~~~ v_{\perp} \gg {\omega_{Be}^{(\rho)}v_z\over
\omega_{Be}^{(z)}} ~.
$$
All terms (see (A8) in Appendix) can be expanded as a power series in $a$.
Integrating over the region
$\tau\rightarrow 0$
(which is most considerable for small $\mid a\mid$), one obtains \\
$\delta f_e = Q(\varphi)\int_0^{\infty} \exp{\{-i\tau Y_1\}}d\tau$,
where
$$
Y_1 = {A_1\over 1+\sin\varphi
+\cos\varphi} + A_1a\cos(\varphi+\varphi_0) + d_1 + 2D -
(d_1+D)a\sin(\varphi+\varphi_0) ~,
$$
i.e. $\delta f_e = -iQ(\varphi)/Y_1 $ . To find the location
of poles, it is assumed, that $S=0$; as a result, we have the
resonance conditions:
$$
{k_zv_z-\omega-v_{\perp}k_{\perp}\sin(\varphi+\varphi_0+\varphi_3+\alpha)\over
\omega_{Be}^{(z)}} =
$$
\begin{equation}
{k_{\perp}v_{\perp}^2\{ \sin(\varphi_3+\alpha) -
2\cos(\varphi_3+\alpha)[1+\cos\varphi+\sin\varphi ]\}\over
(1+\cos\varphi+\sin\varphi)\sqrt{\Omega_1^4\rho^2+v_z^2(\omega_{Be}^{(\rho)})^2}}~.
\end{equation}
Thus, the space-dependence of physical quantities leads to
considerable change in the resonance conditions, except
$k_{\perp}=0$ (for which the well known result follows: 
$k_zv_z-\omega=0$). Note, that the resonance conditions depend
on the space variable $\rho$ and angle $\alpha$. Besides,
the conditions (22) are nonlinear with respect to $v_{\perp}, \varphi ,
v_z$, i.e. in the general case two of these variables cannot be
chosen arbitrarily. Therefore, the resonance region decreases.

In spite of the existence of poles in explicit form for $\mid a
\mid = 1$ and $b_2=1$, the latter conditions are additional; 
therefore, the resonances don't contribute to the Landau damping
after integration in (A9) (see Appendix).

Now we consider the case $\mid a\mid > 1$.
One can easily see from the structure of term $Q(\varphi - \tau)$,
that there are the poles of the integrand at $a > 1~$  for
\begin{equation}
\varphi - \tau + \varphi_{0} = \arcsin(-{1\over a}).
\end{equation}
However, for the additional Landau damping the
oscillations of the exponent in (A8) (see Appendix) near these poles should
be studied. For $a > 1$ the
oscillations do not increase for $\varphi + \varphi_{0} = -\arcsin(1/a)$ and
$\tau = 2n\pi$, $n=0,1,2,...$. However, expanding
$Y(\varphi,\tau)$ in $\tau$ near $\tau=2n\pi$ for $n=1,2,...$,
one can see that the integration in the limits from $-\infty$ to $+\infty$ in
(A8) gives zero. It follows for $n=0$, that
$\delta f_e = Q(\varphi)\int_0^{\infty}\exp{\{ -i\tau Y_2\}}d\tau $,
where
$$
Y_2 = {A_1\over 1 + \sin\varphi
+ \cos\varphi} + {A_1a\cos(\varphi + \varphi_0)\over 1 +
a\sin(\varphi + \varphi_0)} + D -
{(d_1+D)\cos\varphi_1\over \sqrt{a^2-1}(\sin(\varphi+\varphi_0)+\sin\varphi_1)} ~.
$$
Therefore, $\delta f_e = -iQ(\varphi)/Y_2$. The third term in
$Y_2$ is negligible near $\varphi + \varphi_0 =
-\arcsin(1/a)$; the first term is also negligible, except
$\varphi = \pi , 3\pi /2$. We suppose $S=0$ and, using the
definition of $\varphi_1$, obtain the resonance
conditions:
$$
k_zv_z + k_{\perp}v_{\perp}\sin(\varphi_3 + \alpha +
\varphi_1) - \omega = 0 .
$$
Since $\varphi$ is fixed, the resonance doesn't contribute to
the Landau damping. The expression for $Y_2$ can be used for
finding approximate location of poles in the case of $\mid a \mid \gg
1$. The latter resonance gives the Landau damping. The
resonance conditions are:
\begin{equation}
k_zv_z + k_{\perp}v_{\perp}\sin(\varphi_3 + \alpha -\varphi
- \varphi_0) - \omega = -{k_{\perp}v_{\perp}\sin(\alpha +
\varphi_3)\sin(\varphi + \varphi_0)\over 1 + \cos\varphi +
\sin\varphi} ~.
\end{equation}
We can see the transformation of the resonance conditions, which is
a consequence of plasma motion and inhomogeneity.
Note, that (24) gives rise to the well known condition
$k_zv_z-\omega=0$ for $k_{\perp}=0$.
Thus, in the general case there exist resonance particles
(poles in $\delta f_e$), which contribute to the Landau
damping.

We make some remarks on the local-rotational description of
plasma. One can imagine that in a variety of cases the same system is
composed of
different subsystems. The question of the scale of averaging is not trivial.
Since plasma characteristics depend on the regime in it, it is
clear that the averaging over different local subsystems can lead
to different results.
Here the question arises, what particular averaging is correct? 
It is possible that this case just represents the
local-rotational regime (since
there is additional damping as waves propagate in an accelerated medium).
The basic ideas of the description consist in the following.
Instead of using averaged plasma characteristics
for finding some physical quantity, we shall seek the quantity in
a small rotating subsystem with definite parameters
(any plasma particles are involved in the local
rotational movements). The net result (the specific physical
quantity under investigation in a plasma system as a whole) can
be obtained by averaging the quantity over all possible
directions of subsystem rotation.
Actually, we can imagine that the plasma (even though it is collisionless) consists
of rotating parts with radius $r \sim r_{D}$ and linear velocity
$v \sim v_{m}$, where $r_{D}$ is the Debye radius
\begin{equation}
{1\over r_{D}} = \sqrt{{4\pi e^{2}\over T}\sum_{\alpha}K_{\alpha}N_{\alpha}^{2}}~~,
\end{equation}
$v_{m}$ is the most probable velocity
\begin{equation}
v_{m} = \sqrt{{2T\over m}}~~.
\end{equation}

Therefore, the characteristic frequency of the local rotation is
\begin{equation}
\Omega_{l} \sim {v_{m}\over r_{D}}
\end{equation}
(collisions produce local rotations with $r \leq {l_{0}\over 2}$ , where $l_{0}$
is the free path length).

In the general case the solution (but not equations) for a
physical quantity under study (transport coefficients, for
example) should be averaged over the sequence of local
rotating "cells". For a "cell" the
kinetic equation takes the form (for electrons):
$$
{\partial f_{e}\over \partial t} +
\vec{v}{\partial f_{e}\over \partial\vec{r}} +
(-e\vec{E} - {e\over c}[\vec{v}\times\vec{B}] +
2m[\vec{v}\times\vec{\Omega}_l] -
$$
\begin{equation}
{e\Omega_l\rho\over
c}[B_z\vec{e}_{\rho} - B_{\rho}\vec{e}_z] +
m\Omega_{l}^{2}\vec{\rho}){\partial f_{e}\over
\partial\vec{p}} = {f_{e} - f_{0e}\over \tau}~~.
\end{equation}

This way of inclusion of local regimes (with subsequent
averaging over all directions of $\vec{\Omega}_l$)
can be adaptable to the strong turbulence.

\section{THE SELF-SUSTAINED ROTATION}

First of all, we note that any rotation induces
some additional electric field $\vec{E}_0$, which is
directed towards the axis of rotation (z-axis).
Assume that the initially non-rotating plasma system resides in the external magnetic field $\vec{B}_{0} = \vec{e}_{z}B_{0}$. \\
1) If the electron surplus exists ($\Delta > 0$), then
the radial field $\vec{E}_{0\rho}$ is directed to the
$z$-axis. The $\vec{E}_{o\rho}\times\vec{B}$ drift of charged
particles results in the movement round a circle. The
originated centrifugal force causes partial separation
of charges. Therefore, the field $\mid\vec{E}_{0\rho}\mid$ will
increase resulting in sustaining of rotation. Thus,
a collective rotation with $\vec{\Omega}$ directed as
$\vec{B}$ (parallel) arises spontaneously in a plasma system
with an electron surplus.\\
2) If the ion surplus exists (i.e. $\Delta < 0$), then
the radial field $\vec{E}_{0\rho}$ is directed from the
$z$ axis. The $\vec{E}_{0\rho}\times\vec{B}$ drift initiates
rotation with $\vec{\Omega}$ being antiparallel to $\vec{B}$.
However, the partial separation of charges, which is caused by
a centrifugal force, results in decreasing $\mid~E_{0\rho}\mid$.
This effect counteracts the rotation. Since
there exists the rotation with $\vec{\Omega}$ antiparallel to
$\vec{B}$ in a positive (only) particle system (it is some
limit), there exists some critical value of the ion
surplus $\Delta_c$ ($\Delta = \Delta_c < 0$), starting from
which the spontaneous rotation of plasma ($\vec{\Omega}$
antiparallel to $\vec{B}$) is possible.\\
We shall show that the spontaneous rotation with $\vec{\Omega}$
parallel to $\vec{B}$ can arise in an initially neutral plasma
system ($\Delta = 0$). The condition, under which the self-rotation
arises, is $\Omega R_0 < v_{d}$; here $v_d$ is the
drift velocity, i.e. $B_0\Omega R_0/c < -E_{0\rho}$. The
equality in the condition is possible for $\Omega = 0$ and
$\Omega = \Omega_{self}$. Which equilibrium is stable? We note,
that the equiprobable rotational small disturbances
($\vec{\Omega}\rightarrow 0$; or, factually, the ratio of the
rotational energy to $kT$ is small) with $\vec{\Omega}$
parallel and antiparallel to $\vec{B}$ are not equivalent for
the plasma system as a whole. As a result, plasma will rotate
on the average in some definite direction: any rotational
disturbance produces the radial field $\vec{E}_{0\rho}$
directed toward the axis of rotation, but the appeared
$\vec{E}_{o\rho}\times\vec{B}$ drift promotes rotational
disturbances with $\vec{\Omega}$ parallel to $\vec{B}$ only.
We have for the other limit (the case of externally initiated large
$\vec{\Omega}$, i.e. $M_i\Omega^2\rho^2/(2kT) \gg 1$) the inequality
$\Omega\rho_0 > v_d$, since the limitations on $E_{0\rho}$
always exist for any plasma system: $-\infty <
E_{0\rho}^{min} \le E_{0\rho} \le E_{0\rho}^{max} < \infty$.
Therefore, the drift opposes the
rotation. Thus, the rotation tends to some equilibrium
$\Omega_{self} \ne 0$ for $\Delta = 0$.

In the general case, when the rotation ($\vec{\Omega}$ is parallel to $\vec{B}_{0}$)
arises by any reason, the separation of charges and
electric field arise. As a result, the $\vec{E}\times\vec{B}$ drift
of charged particles (the movement round a circle) arises. This movement will
either add to the initial rotation or
subtract from it. Eventually, the stable rotation is established. From
the equation
$$
v_{d} = \Omega(\rho, z)\rho,
$$
where the drift velocity is
$$
\vec{v}_{d} = -{cE_{0\rho}(\rho,z)\vec{e}_{\varphi}\over B_{0}}
$$
it follows, that the stable rotation frequency is
\begin{equation}
\Omega_{self}(\rho, z) = -{cE_{0\rho}(\rho, z)\over \rho B_{0}},
\end{equation}
with $E_{0\rho}(\rho, z)$ taken from (6) and $\mid B_z \mid \ll
\mid B_0 \mid$.

Under an assumption, that the originated magnetic field
$\vec{B}_{z}$ is comparable
to $\vec{B}_{0}$, the following substitution needs to be done in equation (25):
$B_{0} \rightarrow  B_{0} + B_{z} $ with
$B_{z}$ taken from (10).
In this case the self-sustained rotation of plasma takes place.

We make some remarks.\\
1) Indeed, to evaluate the rotational state stability
one can involve the energy balance. In this case the energy
surplus characterizes qualitatively the rate (or time) of transition.
However, a rather complex integral system of equations for $\Omega_{self}$ 
can be solved numerically (or by a perturbative method) 
for all specific values of
plasma system parameters only.\\
2) $\Omega_{self}$ tends to zero for the temperature high enough.\\
3) If the external magnetic field $\vec{B}_0$ is zero, then
the self-sustained rotation is impossible (the velocity
$\rho\Omega_{self}$ must be greater than the speed of
light).\\
4) If several quasistable field distributions (distinguished by a potential 
barrier) can exist, then the existence of several quasistable rotating regimes
(with $\Omega_{self}^j$) is possible.

\section{CONCLUSIONS}

The generation of electric and magnetic nonhomogeneous self-fields in rotating plasma
can be explained by various actions of the centrifugal force on dissimilar
particles. There are different $\rho$-distributions of these particles
and the partial separation of charges. Therefore, the
electric field arises (see (6),(8)).
The magnetic field (see (9),(10)) exists as a result of rotation of charged regions
(the magnetic actions are not compensated).  As for  relation  to
the physics  topics  from  Sec.I,  there  are  new  peculiarities
regarding the  self-consistent   description.   All   self-fields
$E_{\rho}(\rho,z), E_{z}(\rho,z),               B_{\rho}(\rho,z),
B_{z}(\rho,z)$ are taken into consideration. The self-sustained
rotation (which origins from initially neutral state)
exists as a result of the particles drifts in the electric
self-field and external magnetic field (see (29)). This is a 
new result. The dielectric constant
tensor in rotating plasma differs from that in non-rotating
plasma. As the result of plasma inhomogeneity and accelerated motion, 
the resonance conditions change (see (22),(24)).
There are new poles in a disturbed distribution function
and, therefore, new terms in the Landau damping. The
local-rotational description (over rotating "cells") of plasma can be useful.
\vskip 0.3cm
\centerline{\large{{\bf APPENDIX: Derivation of the disturbed}}}
\centerline{\large{{\bf distribution function}}}
\vskip 0.3cm

The problem is to obtain the disturbed electron
distribution function.
We substitute
(17-21) and the force (13)
into the kinetic equation (16), correct to the second order ($\delta f, \vec{E'}, \vec{B'}$
are small). Then it turns out that
$$
i(\vec{k}'\vec{v} - \omega)\delta f_{e} + {ef_{0e}\over T}(\vec{v}\vec{E}') -
{ef_{0e}\over Tc}\Omega\rho (B_{\rho}'v_{z} - B_{z}'v_{\rho}) +
$$
$$
m\Omega_{1}^{2}(\rho,z)\vec{\rho}{\partial\delta f_{e}\over \partial\vec{p}} -
{e\over c}[\vec{v}\times\vec{B}_{01}]{\partial\delta f_{e}\over \partial\vec{p}} -
e\vec{E}_{0z}^{(1)}{\partial\delta f_{e}\over  \partial\vec{p}} =
0 , \eqno{(A1)}
$$
where
$$
\Omega_{1}^{2}(\rho,z) = \Omega^{2} -
{eE_{0\rho}(\rho,z)\over m\rho} - {eB_{0z}\over mc}\Omega~,
~~\vec{B}_{01}=\vec{B}_0-{2mc\over e}\vec{\Omega}~,
$$
$$
\vec{E}_{0z}^{(1)}=\vec{E}_{0z}-{\Omega\rho\over
c}B_{0\rho}\vec{e}_z~ .
$$
Transferring to the cylindrical coordinate system $(v_{z}, v_{\perp}, \varphi)$ in the
velocity space (Fig. 2), one gets
\begin{figure}
\unitlength=1mm
\special{em:linewidth 0.4pt}
\linethickness{0.4pt}
\begin{picture}(115.00,80.00)
\put(25.00,40.00){\vector(1,0){30}}
\put(55.00,40.00){\vector(1,1){22}}
\put(55.00,40.00){\vector(1,1){10}}
\put(55.00,40.00){\vector(1,-1){20}}
\put(77.00,62.00){\vector(-1,1){10}}
\put(25.00,35.00){$0$}\put(51.00,35.00){$\vec{\rho}$}
\put(76.00,34.00){$\alpha$}\put(70.00,60.00){$\vec{v}_{\perp}$}
\put(70.00,17.00){$\vec{k}_{\perp}$}\put(70.00,45.00){$\varphi$}
\put(57.50,49.50){$\vec{e}_{v_{\perp}}$}\put(65.00,67.00){$\vec{e}_{\varphi}$}
\put(7.00,5.00){FIG.2. The cylindrical coordinate system in the velocity space.}
\multiput(55.00,40.00)(5.00,0.00){6}{\ldots}
\qbezier(65.00,30.00)(74.00,40.00)(65.00,50.00)
\qbezier(69.00,26.00)(74.00,29.50)(75.00,40.00)
\end{picture}
\end{figure}
$$
i[k_{z}v_{z} + k_{\perp}v_{\perp}\cos\varphi - \omega +
S(\vec{r})\rho v_{\perp}\cos(\varphi - \alpha) +
S(\vec{r})zv_{z}]\delta f_{e} + {ef_{0e}\over T}(\vec{v}\vec{E}') -
$$
$$
{ef_{0e}\Omega\rho\over Tc}(B_{\rho}'v_{z} - B_{z}'v_{\perp}\cos{(\varphi - \alpha)}) +
\{ \Omega_{1}^{2}\rho\cos(\varphi - \alpha) - \omega_{Be}^{\rho}\sin{(\varphi - \alpha)}v_z\} {\partial\delta f_{e}\over \partial v_{\perp}} +
$$
$$
[\omega_{Be}^{(z)} - {\Omega_{1}^{2}\rho\over
v_{\perp}}\sin(\varphi - \alpha) -
\omega_{Be}^{\rho}\cos{(\varphi - \alpha)}{v_z\over
v_{\perp}}]{\partial\delta f_{e}\over \partial\varphi} +
$$
$$
[\omega_{Be}^{\rho}\sin{(\varphi -     \alpha    )}v_{\perp}    -
{eE_{0z1}^{(1)}\over   m}]{\partial\delta   f_{e}\over   \partial
v_{z}} = 0~ , \eqno{(A2)}
$$
where $k_{\perp}$ is the given value, $\sin\alpha = {d\over \rho}$, $d$ is the minimum  distance from
the center to the ray. Then we obtain the equation
$$
{\partial\delta f_{e}\over \partial\varphi} + i{a_{1}\sin\varphi + b_{1}\cos\varphi + d_{1}\over
a_{2}\sin\varphi + b_{2}\cos\varphi + 1}\delta f_{e} = Q(\varphi)
, \eqno{(A3)}
$$
where
$$
a_{1} = {S(\vec{r})v_{\perp}d\over \omega_{Be}^{(z)}}~~ ,~~ b_{1} = {k_{\perp}v_{\perp} +
S(\vec{r})\sqrt{\rho^{2} - d^{2}}v_{\perp}\over \omega_{Be}^{(z)}}~ ,~
d_{1} = {k_{z}v_{z} - \omega + S(\vec{r})v_{z}z\over \omega_{Be}^{(z)}}~ , ~~
$$
$$
b_{2} = {\Omega_{1}^{2}\rho\over
v_{\perp}\omega_{Be}^{(z)}}\sin{\alpha} - {\omega_{Be}^{(\rho
)}v_z\over \omega_{Be}^{(z)}v_{\perp}}\cos{\alpha}~ , ~~
a_{2} = -{\Omega_{1}^{2}\rho\over
v_{\perp}\omega_{Be}^{(z)}}\cos{\alpha} - {\omega_{Be}^{(\rho
)}v_z\over \omega_{Be}^{(z)}v_{\perp}}\sin{\alpha}~,
$$
$$
Q(\varphi) = {W\over \omega_{Be}^{(z)}(a_{2}\sin\varphi + b_{2}\cos\varphi + 1)}~ , ~~~
W = -{ef_{0e}\over T}\{ (\vec{v}\vec{E}') - {\Omega\rho\over c}[B_{\rho}'v_{z} -
$$
$$
B_{z}'v_{\perp}\cos{(\varphi - \alpha )}]\} +
[{eE_{0z1}^{(1)}\over m} - \omega_{Be}^{(\rho)}\sin{(\varphi -
\alpha )}v_{\perp}]{\partial\delta f_{e}\over \partial v_{z}} -
$$
$$
[\Omega_{1}^{2}\rho\cos{(\varphi - \alpha )} -
\omega_{Be}^{(\rho)}v_z\sin{(\varphi - \alpha )}]{\partial\delta f\over
\partial v_{\perp}} ~.
$$

The solution of equation (A3) is
$$
\delta f_{e}         =         e^{-iy(\varphi)}\int_{C}^{\varphi}
e^{iy(\varphi')}Q(\varphi')d\varphi' ,\eqno{(A4)}
$$
where
$$
{dy\over d\varphi}  =  {a_{1}\sin\varphi  +  b_{1}\cos\varphi   +
d_{1}\over a_{2}\sin\varphi + b_{2}\cos\varphi + 1} , \eqno{(A5)}
$$
that is,
$$
y(\varphi) = {b_{1}a_{2} - a_{1}b_{2}\over a_{2}^{2} + b_{2}^{2}}
\ln\Biggl | {a_{2}\sin\varphi + b_{2}\cos\varphi + 1\over 1 +
\tan{{\varphi\over 2}}}\Biggr | +
$$
$$
{b_{1}b_{2} + a_{1}a_{2}\over  a_{2}^{2}  +  b_{2}^{2}}\varphi  +
\Big   (  d_{1}  -  {b_{1}b_{2}  +  a_{1}a_{2}\over  a_{2}^{2}  +
b_{2}^{2}}\Big  )  \int   {d\varphi'\over   a_{2}\sin\varphi'   +
b_{2}\cos\varphi' + 1} ,\eqno{(A6)}
$$
$$
 \int {d\varphi'\over a_{2}\sin\varphi' + b_{2}\cos\varphi' + 1} =
$$
$$
1)~~~~~~~~~~~~~~~~~ {2\over \sqrt{1- b_{2}^{2} - a_{2}^{2}}}\arctan{(1 - b_{2})\tan{{\varphi\over 2}} + a_{2}\over \sqrt{1 - b_{2}^{2} - a_{2}^{2}}} ~, ~~ b_{2}^{2} + a_{2}^{2} < 1 ,
$$
$$
2)~~~~~~ {1\over \sqrt{b_{2}^{2} + a_{2}^{2} - 1}}\ln\Biggl | {(1 - b_{2})\tan{\varphi\over 2} +
a_{2} - \sqrt{b_{2}^{2} + a_{2}^{2} - 1}\over (1- b_{2})\tan{\varphi\over 2} + a_{2} + \sqrt{b_{2}^{2} + a_{2}^{2} -1}}\Biggr | , ~~ b_{2}^{2} + a_{2}^{2} > 1 ,
$$
$$
3)~~~~~~~~~~~~~~~~~~~~~~~~~~ {1\over a_{2}}\ln\Biggl | 1 + a_{2}\tan{\varphi\over 2}\Biggr | ,~~ b_{2} = 1 ,
$$
$$
4)~~~~~~~~~~~~~~~~~~~~~~~ -{2\over a_{2} + (1 - b_{2})\tan{\varphi\over 2}} ,~~ a_{2}^{2} + b_{2}^{2} = 1 .
$$
Under the $\varphi$-periodicity condition on $\delta f_e$ it
turns out from (A4) (with the substitution $\varphi'=\varphi
-\tau$), that this expression (note that all functions are
$\varphi$-periodic) can possess the
$\varphi$-periodicity for $C=\pm\infty$;~ the value $C=\infty$
obeys all transitions (in limits) to the well-known results only; therefore,
$$
\delta f_{e}  =  \int_{0}^{\infty}  e^{i[y(\varphi  -   \tau)   -
y(\varphi)]}Q(\varphi - \tau)d\tau .\eqno{(A7)}
$$
The substitutions
$$
\varphi_{0} = \arctan{b_{2}\over a_{2}} ,
~~~A_1 = -{v_{\perp}[k_{\perp}\sin{(\varphi_3 +\alpha)} +
\rho S\sin{\varphi_3}]\over a\omega_{Be}^{(z)}} ,~~
\tan{\varphi_3} = {\Omega_1^2\rho\over
\omega_{Be}^{(\rho)}v_z}~ ,
$$
$$
a^2 = {\Omega_{1}^{4}\rho^2\over
v_{\perp}^2(\omega_{Be}^{(z)})^2} +
{(\omega_{Be}^{(\rho)})^2v_z^2\over (\omega_{Be}^{(z)})^2v_{\perp}^2}~ ,~~
D = {v_{\perp}[k_{\perp}\cos{(\alpha + \varphi_3)} +
\rho S\cos{\varphi_3}]\over a\omega_{Be}^{(z)}}~ ,
$$
$C_{1} = d_1 + D~ ,~~ \varphi_1 = \arctan{(1/\sqrt{a^2-1})}$\\
and some rearrangements give:
$$
\delta f_{e} = \int_{0}^{\infty} \exp{\{ iY(\varphi , \tau)\} }
Q(\varphi - \tau)d\tau~ ,\eqno{(A8)}
$$
$$
Y(\varphi , \tau) = A_1\ln\Biggl | {[1 + a\sin{(\varphi + \varphi_{0} - \tau)}]\over
[1 + a\sin(\varphi + \varphi_{0})]}{[1 + \tan{{\varphi\over
2}}]\over [1 + \tan{({\varphi - \tau\over 2})}]}\Biggr | -
D\tau + C_{1}W_{1} ~,
$$
where $W_{1} = $
$$
1)~~~~~~~~~~ -{2\over \sqrt{1 - a^{2}}}\arctan{\Biggl [ {\sqrt{1 - a^{2}}\sin{\tau\over 2}\over
\cos{\tau\over 2} + a\sin(\varphi + \varphi_{0} - {\tau\over
2})}\Biggr ]} , ~~\mid a\mid < 1 ~,
$$
$$
2)~~~~ {1\over \sqrt{a^{2} - 1}}\ln{\Biggl | {\sin{({\varphi +
\varphi_0 + \varphi_1 - \tau\over 2})}\cos{({\varphi +
\varphi_0 - \varphi_1\over 2})}\over \sin{({\varphi +
\varphi_0 + \varphi_1\over 2})}\cos{({\varphi + \varphi_0 -
\varphi_1 - \tau\over 2})}}\Biggr | }~ ,~~ \mid a\mid ~> 1 ,
$$
$$
3)~~~~~~~~~~~~~~~~~~~~~~ {1\over a_{2}}\ln{\Biggl | {1 + a_{2}\tan{{\varphi - \tau\over 2}}\over 1 +
a_{2}\tan{\varphi\over 2}}\Biggr | } ,~~ b_{2} = 1 ,
$$
$$
4)~~~~~~~~~~~~~~~ -{2\sin{\tau\over 2}\over \cos{\tau\over 2} + \sin(\varphi + \varphi_{0} -
{\tau\over 2})} ,~~ \mid a\mid = 1 ,
$$
$$
Q(\varphi - \tau)={- 1\over 1+a\sin(\varphi-\tau+\varphi_{0})}\Biggl [ {ef_{0e}\over
\omega_{Be}T}[ (\vec{v}\vec{E}") - {\Omega\rho\over c}(B_{\rho}"v_{z} 
$$
$$
-
 B_{z}'v_{\perp}\cos{(\varphi-\alpha-\tau)})]+
av_{\perp}\sin(\varphi_3 + \alpha - \varphi + \tau){\partial\delta f_{e}\over \partial v_{\perp}} 
$$
$$
-
 \biggl ( {eE_{0z1}^{(1)}\over m} -
\omega_{Be}^{(\rho)}v_{\perp}\sin(\varphi - \alpha -
\tau)\biggr ){\partial\delta f_{e}\over \partial v_{z}}\Biggr ],
$$
where the vectors $\vec{E}" , \vec{B}"$ are the vectors $\vec{E}' , \vec{B}'$
turned at angle $\tau$ in the plane normal to $\vec{B}_{0}$.

Note that expression (A8) is rigorous. As the first approximation, one can use 
$$
{\partial\delta f(\varphi - \tau)\over \partial v_{\perp}} =
{\partial\delta f_{M}(\varphi - \tau)\over \partial v_{\perp}}~ ,~~~ {\partial\delta f(\varphi - \tau)\over \partial v_{z}} =
{\partial\delta f_{M}(\varphi - \tau)\over \partial v_{z}},
$$
where the disturbed part of the Maxwell velocity distribution
function $^{18}$ is
$$
\delta f_{M}(\varphi) = \int_{0}^{\infty} \exp{(-i\alpha_1\tau' - 2i\beta\cos(\varphi -
{\tau'\over 2})\sin{\tau'\over 2})}Q^{*}(\varphi - \tau')d\tau' ,
$$
$$
\alpha_1 = {k_{z}v_{z} - \omega\over \omega_{Be}^{(z)}} , ~~~ \beta = {k_{\perp}v_{\perp}\over
\omega_{Be}^{(z)}} ,~~ ~ Q^{*}(\varphi) = -{e(\vec{v}\vec{E}')f_{M}\over T\omega_{Be}^{(z)}} .
$$
Here $f_{M}$ is the Maxwell velocity distribution function.
Note from (A8), that there exist different regimes of
particle behaviour (they depend on the velocity of 
particles). The dielectric constant tensor can be found from
$$
{\varepsilon_{\alpha\beta} - \delta_{\alpha\beta}\over 4\pi}E_{\beta} =
{e\over
i\omega}\int_{0}^{\infty}\int_{0}^{2\pi}\int_{-\infty}^{+\infty}
v_{\alpha}\delta f_{e}dv_{z}d\varphi         v_{\perp}dv_{\perp}~
.\eqno{(A9)}
$$
The dielectric constant tensor
differs from that in non-rotating plasma.

{}

\end{document}